# Liquid Phase 3D Printing for Quickly Manufacturing Metal Objects with Low Melting Point Alloy Ink

**Lei Wang [1], Jing Liu [1,2]***

**1.** Beijing Key Lab of CryoBiomedical Engineering and Key Lab of Cryogenics,

Technical Institute of Physics and Chemistry,

Chinese Academy of Sciences, Beijing, China

**2.** Department of Biomedical Engineering, School of Medicine,

Tsinghua University, Beijing, China

**\*Address for correspondence:**

Dr. Jing Liu

Beijing Key Lab of CryoBiomedical Engineering,

Technical Institute of Physics and Chemistry,

Chinese Academy of Sciences,

Beijing 100190, China

E-mail address: jliu@mail.ipc.ac.cn

Tel. +86-10-82543765

Fax: +86-10-82543767

**Abstract**

Conventional 3D printings are generally time-consuming and printable metal inks are rather limited. From an alternative way, we proposed a liquid phase 3D printing for quickly making metal objects. Through introducing metal alloys whose melting point is slightly above room temperature as printing inks, several representative structures spanning from one, two and three dimension to more complex patterns were demonstrated to be quickly fabricated. Compared with the air cooling in a conventional 3D printing, the liquid-phase-manufacturing offers a much higher cooling rate and thus significantly improves the speed in fabricating metal objects. This unique strategy also efficiently prevents the liquid metal inks from air oxidation which is hard to avoid otherwise in an ordinary 3D printing. Several key physical factors (like properties of the cooling fluid, injection speed and needle diameter, types and properties of the printing ink, etc.) were disclosed which would evidently affect the printing quality. In addition, a basic route to make future liquid phase 3D printer incorporated with both syringe pump and needle arrays was also suggested. The liquid phase 3D printing method, which owns potential values not available in a conventional modality, opens an efficient way for quickly making metal objects in the coming time.

**Keywords:** 3D printing; Liquid phase manufacture; Liquid metal printer; Rapid prototyping; Low melting point metal; Liquid cooling; Oxidation prevention
1. Introduction

The rapid prototyping (RP) is a kind of additive manufacturing (AM) technology which is found increasingly important in a variety of newly emerging areas including chemical synthesis,[1]

2
**Abstract**

Conventional 3D printings are generally time-consuming and printable metal inks are rather limited. From an alternative way, we proposed a liquid phase 3D printing for quickly making metal objects. Through introducing metal alloys whose melting point is slightly above room temperature as printing inks, several representative structures spanning from one, two and three dimension to more complex patterns were demonstrated to be quickly fabricated. Compared with the air cooling in a conventional 3D printing, the liquid-phase-manufacturing offers a much higher cooling rate and thus significantly improves the speed in fabricating metal objects. This unique strategy also efficiently prevents the liquid metal inks from air oxidation which is hard to avoid otherwise in an ordinary 3D printing. Several key physical factors (like properties of the cooling fluid, injection speed and needle diameter, types and properties of the printing ink, etc.) were disclosed which would evidently affect the printing quality. In addition, a basic route to make future liquid phase 3D printer incorporated with both syringe pump and needle arrays was also suggested. The liquid phase 3D printing method, which owns potential values not available in a conventional modality, opens an efficient way for quickly making metal objects in the coming time.

**Keywords:** 3D printing; Liquid phase manufacture; Liquid metal printer; Rapid prototyping; Low melting point metal; Liquid cooling; Oxidation prevention

1. **Introduction**

The rapid prototyping (RP) is a kind of additive manufacturing (AM) technology which is found increasingly important in a variety of newly emerging areas including chemical synthesis,[1] microfluidics,[2] tissue engineering,[3-5] electronic circuit and device,[6-7] etc. The basic principle of RP technology is to create a three dimensional object through laying down successive layers of materials which can be powdered plastic, metal particles or any other adhesive materials. So far, the AM techniques for metals have been developed for more than 20 years. Among them, three typical ways including laser sintering (LS), laser melting (LM) and laser metal deposition (LMD) are the most prevailing ones which are generally capable of processing a variety of high melting point metals, alloys and metal matrix composites (MMCs).[8] To achieve favorable metal objects during these fabrications, one has to select both appropriate powder materials (whose properties include chemical constituents, particle size, powder flowability, etc.) and laser process (e.g. laser type and power, scan speed, powder layer thickness, etc.).[8-11] For this reason, the currently available types of printable metal inks are rather limited if one wishes to use this conventional 3D metal printing method.

In recent years, the low melting point liquid metal, especially room temperature liquid metal kept attracting more and more extensive attentions in the areas of computer chip cooling, thermal interface material, microfluidics and so on.[12-17] Such material has also been proposed as printing ink with evident values in direct writing electronics and 3D printing technology. Zheng et al initiated a desktop printing of flexible circuits on paper via developing liquid metal ink and established a basic 3D printing scheme including computer controlled machine system for simultaneously manufacturing mechanical structure as well as conductive functional devices.[18] Later, Ladd et al tested a simple way of directly applying syringe to realize liquid metal objects at room temperature.[19] Yu et al discovered and proposed a channelless fabrication method for large-scale preparation of room temperature liquid metal droplets.[20] In these studies, the eutectic alloy of gallium and indium (melting point ~15.7 ℃) was adopted as the writing ink or raw material. A limitation of such room temperature metal inks is that the printed objects are easily subjected to melt which therefore may restrict the application of the device to some extent.

In this study, to further improve the 3D printing techniques for fabricating metal objects, an alternative approach which differs from the existing air cooled 3D printing is proposed for the first time, which can be termed as liquid phase 3D printing. For illustration purpose, the metals whose melting points are above room temperature (20-30 ℃) and less than 300 ℃ were identified and adopted as the printing ink. This is different from the traditional metal printing method in which the high melting point metal inks are often used.

To ensure the printing quality, some basic fluid mechanics issues such as droplet formation in liquid-liquid systems, dripping-jetting transition and so on were systematically studied. Over the years, there have been a lot of experimental and dynamic researches on the principle of the drop formation at a capillary tip.[21-23] When the liquid metal is injected into another immiscible fluid via the needle, two drop formation mechanisms will be observed. If the injection velocity of the liquid metal is lower than a certain critical value, the drops will be formed directly at the needle tip. But if the injection velocity is larger than the critical value, the liquid metal will form a jet which then breaks up into droplets because of Rayleigh instabilities.[21] Generally speaking, in the former case, the droplet size is determined by the buoyancy, viscosity, surface tension and inertial of the fluid and the drop, while in the latter the droplet size is determined by the jet stability dynamics.[22] Theoretical analysis and numerical simulations have been carried out to describe qualitatively and quantitatively the features of the droplet detachment from a needle's tip.[23-25] The present study also raised interesting scientific issues such as the fluids interactions between dropping liquid metal and the base cooling fluid as well as the practical strategies to precisely control the deposition quality of the final metal objects in liquid phase.

**2. Experimental**

2.1 Preparation of Printable Metal Inks

For the present liquid phase 3D printing, all the pure metals or alloys whose melting points are from around room temperature to 300 ℃ can possibly be adopted as printing inks. These include gallium-, bismuth- and indium-based alloys. Addition of nanoparticles such as copper, silver particles into such metal fluids also offers a method to fabricate functional inks as desired. Besides, combination of metal and nonmetal material together can be used to make diverse printing inks. Here, as the first trial along this direction, the $Bi_{35}In_{48.6}Sn_{15.9}Zn_{0.4}$ alloy is specifically selected as printing ink to demonstrate the basic working principle of liquid phase 3D printing method. The preparation process for making this kind of functional ink was as follows: four metals of bismuth, indium, tin and zinc (with high purity of 99.99%) are weighed according to the ratio of 35:48.6:15.9:0.4. These pure metals are put in a beaker for 5 hours at 245 ℃ in an electric vacuum drying oven. Then, the mixture is stirred in the beaker which is put in water bath at 85-90 ℃ for 30 min. Finally, keep the beaker in the electric vacuum drying oven for 2 hours to further ensure a well-mixed alloy solution.

2.2 Preparation of Liquid Phase Cooling Fluid

The liquid phase cooling fluid can be selected from among water, ethanol, kerosene, gluewater, silicone oil, silica gel and so on. Here for brevity, only water and ethanol are adopted as the cooling fluids in a comparative study.

2.3 Experimental Devices

The experimental device used in this study is illustrated in Fig. 1. Because the melting point of Bi35In48.6Sn15.9Zn0.4 is slightly higher than the room temperature, such liquid metal would easily subject to blocking in the syringe needle due to solidification. To solve the problem, the syringe is installed in an aluminum alloy cylinder which is heated via constantan resistance wire (62 ohms per meter). A temperature controller is used to maintain constant temperature of the metal cylinder by adjusting the supply electrical power to the constantan resistance wire. The nitrogen cylinder is used to provide a constant pressure on the syringe and the pressure is regulated by a solenoid valve. The syringe needle is immersed into the liquid phase cooling fluid which is water/ethanol in this experiment. The dripping or jetting process is monitored with a high speed camera (Nikon NR-S3) which can capture 30 frames per second with the exposure time set to 1.999s.

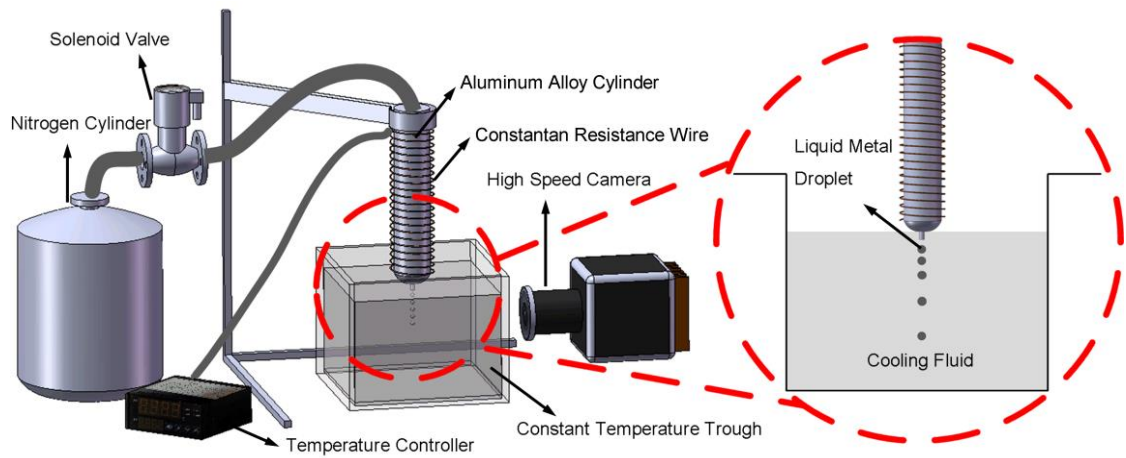

Figure 1 The schematic diagram of the experimental apparatus.

## 3. Results and Discussion

Figure 1 illustrates the basic experimental setup for the present liquid phase 3D printing in making metal objects. Particularly, a four-element alloy $Bi_{35}In_{48.6}Sn_{15.9}Zn_{0.4}$ was developed and adopted here as the printing inks. Some of its properties are measured and provided in Table 1 and Fig. 2. It can be seen that the density of $Bi_{35}In_{48.6}Sn_{15.9}Zn_{0.4}$ (7.564 g/ cm$^3$) is close to that of iron (7.86 g/ cm$^3$). The melting point (58.3 ℃) of this alloy is slightly higher than room temperature and the degree of subcooling is low (2.4 ℃). This means that $Bi_{35}In_{48.6}Sn_{15.9}Zn_{0.4}$ will be cooled quickly while the temperature is reduced in the range of 50-60 ℃. According to our measurements, the melting enthalpy and specific heat capacity of $Bi_{35}In_{48.6}Sn_{15.9}Zn_{0.4}$ (28.94 J/g and 0.262 J/(g*℃), respectively) is much smaller than that of common metals such as iron (272.2 J/g and 0.46 J/(g*℃), respectively) and aluminum (393.0 J/g and 0.88 J/(g*℃), respectively). These behaviors enable its easy liquid-solid phase transition during the printing process. In a word, we identified that $Bi_{35}In_{48.6}Sn_{15.9}Zn_{0.4}$ is an ideal liquid metal printing ink to implement the liquid phase 3D printing as proposed in this paper.

Table 1 Typical physical properties of $Bi_{35}In_{48.6}Sn_{15.9}Zn_{0.4}$

| Density (g/cm$^3$) | Melting point (℃) | Degree of subcooling (℃) | Melting enthalpy (J/g) | Specific heat capacity (J/(g*℃)) |
|---|---|---|---|---|
| 7.564 | 58.3 | 2.4 | 28.94 | 0.262 at 25 ℃ |

Figure 3 presents several typical metal objects with structures spanning from simple to complex dimensions built up by the present liquid phase 3D printing method. For the zero dimensional case, a large number of metal balls can be rapidly formed through dropping the printing ink into the liquid phase (see Fig. 3 (A)) which is a conceptual innovation over existing solder ball manufacturing technology. The size of the metal balls can be adjusted by changing the injection needle diameter, the injection speed, the properties of the cooling fluid and so on. Although the similar preparation method has been mentioned before, it was mainly for air cooling case. Besides the metal balls, one-dimensional linear structure can also be easily manufactured. The liquid metal rods as shown in Fig. 3 (B) were printed along the vertical direction. These structures are somewhat hard to directly make through air cooling or sand cooling method in a conventional 3D printing. Many other structures can also be made in the same way in short time. For example, a frustum of a cone structure and a cylinder structure made of liquid metal are presented in Fig. 3 (C) and Fig. 3 (D), respectively. Overall, the rapid prototyping of these metal objects are attributed to the high thermal conductivity and heat capacity of the liquid phase cooling fluid compared to other cooling mediums. Although the currently available printing resolution is still low, improvements can be very possible through regulating various printing factors in the near future, as will be partially disclosed in later sections. Overall, the liquid phase printing opens the possibility to make complex metal structures via an extremely rapid way.

The formation and deposition processes of the droplets are the core problems to the liquid phase 3D printing method. Fig. 4 presents a sequential process of such droplet formation. It can be seen from Fig. 4 (A) that when the droplet falling velocity is low (3.34 mm/s), the ink tends to form spherical beads under the needle because of the large surface tension between the liquid metal and the cooling fluid. With the increase of the droplet falling speed, adjacent droplets get closer and closer and finally join together. Fig. 4 (B) shows this interesting long tail tadpole-like droplets phenomenon.

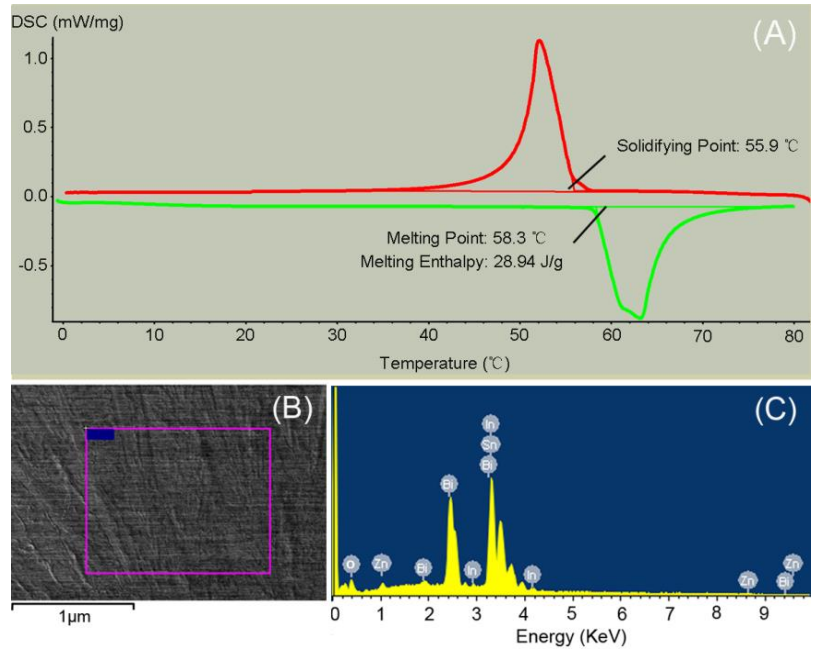

Figure 2 Basic properties of $Bi_{35}In_{48.6}Sn_{15.9}Zn_{0.4}$: (A) Differential scanning calorimetry (DSC) curve; (B) Scanning electron microscopy (SEM) image; (C) Energy dispersive spectrum (EDS).

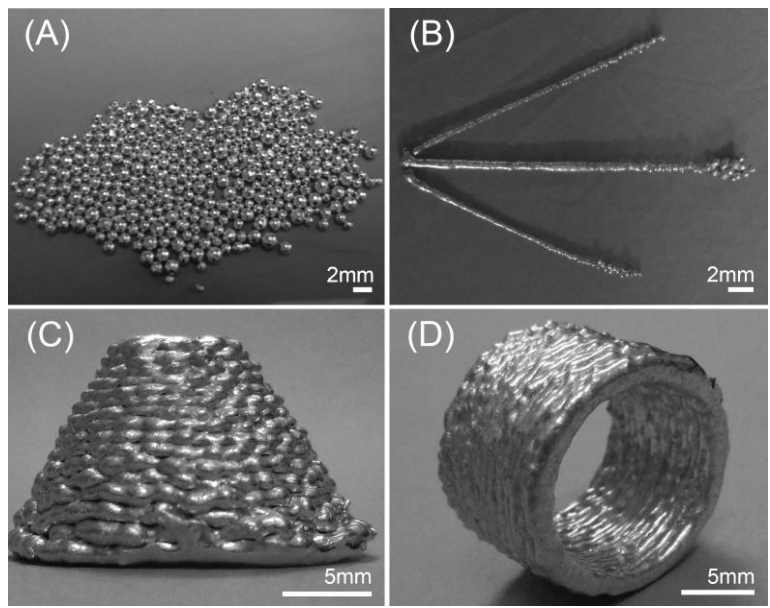

Figure 3 Typical 3D metal structures made by liquid phase 3D printing method. (A) Liquid metal balls; (B) Liquid metal rods; (C) Frustum of a cone structure; and (D) Cylinder structure.

The desired metal objects are formed by deposition of fused droplets. Because the melting point of the printing ink used in this paper is slightly higher than that of room temperature, the droplets are easy to be melt and solidified. When one droplet falls onto the solid column below, the top of the column absorbs the heat and fuses with the droplet. As the droplet is cooled to the temperature of the cooling fluid because of heat transfer, it becomes a part of the column which will then grow into certain structures as desired. This is the basic principle of liquid phase printing whose typical process is shown in Fig. 5.

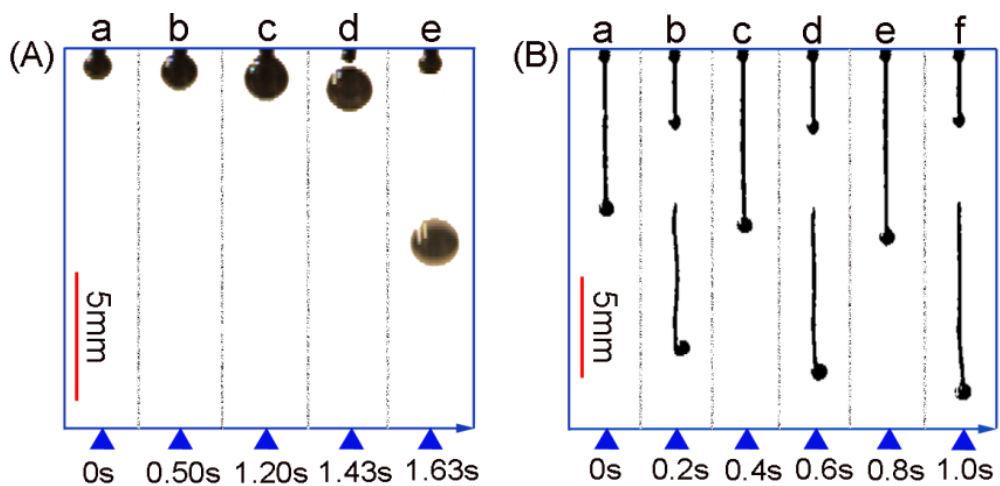

Figure 4 Dynamic droplet configurations in liquid phase: (A) The droplet formation process in water phase (the droplet falling velocity is 3.34 mm/s); (B) The long tail tadpole-like droplets due to fast injection speed in ethanol phase (the falling velocity is 7.98 mm/s)

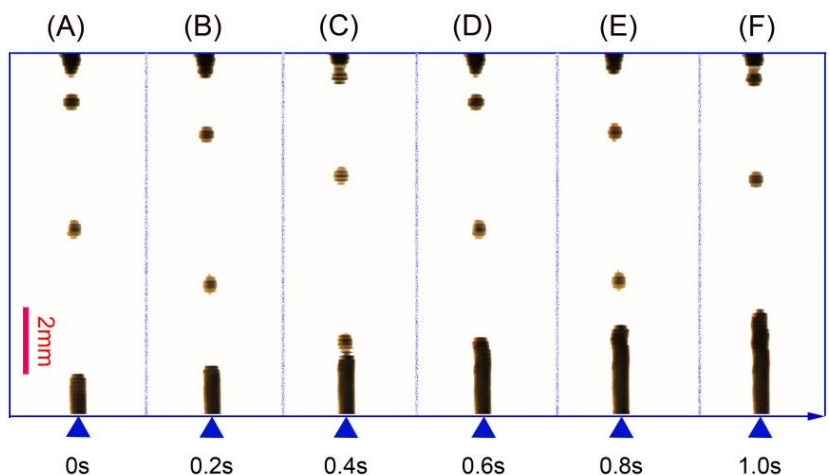

Figure 5 The droplet deposition process (from A to F) in ethanol phase (the droplet falling velocity is 5.65 mm/s)

Figure 6 provides a comparison between ethanol cooling and air cooling printings when the droplets fall down at the same speed and from the same needle. The droplets are in semi-solid state when they contact the bottom as the ethanol cooling method is administrated. This is owing to the rapid heat dissipation of ethanol and a cluster structure can be easily formed which is shown in Fig. 6 (A). But for the air cooling printing which is the conventional 3D printing case, the droplets would keep in melting state within a much longer time when they reach the bottom and a big molten globule is thus formed. In addition, metal oxidation is more severe than that when using the ethanol as the cooling medium (see Fig. 6 (B)).

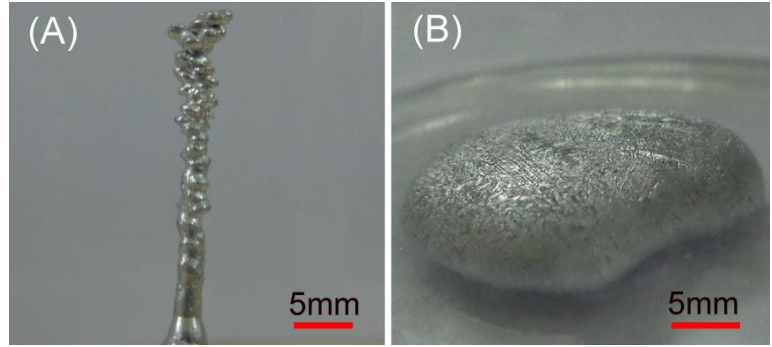

Figure 6 Comparison between ethanol cooling and air cooling printings: (A) Column formed by the ethanol cooling method; (B) Molten globule formed by the air cooling approach.

Table 2 comparatively lists the thermal conductivity, density, heat capacity of water, ethanol and dry air, respectively. It can be found that the relative densities of water and ethanol are 828.22 and 655.02, respectively. According to Archimedes' principle, the upward buoyant force exerted on a body immersed in a liquid is equal to the weight of the amount of fluid it displaces. For a liquid metal droplet immersed in the fluid, its buoyancy can be expressed as

$$F = G_{droplet} = \rho_{droplet} \cdot g \cdot V_{droplet} \tag{1}$$

where, $G_{droplet}$, $\rho_{droplet}$, $V_{droplet}$ are the gravity force, density and volume of the droplet, respectively; $g$ is the gravitational acceleration which equals to 9.80665 m/s$^2$. Therefore the upward buoyancy forces of a droplet immersed in water and ethanol are 828.22 and 655.02 times larger than that of air cooling case. It is the larger buoyancy that contributes to the buffering action. The relative thermal conductivities of water and ethanol are 23.05 and 9.27, respectively. And the relative heat capacities of water and ethanol are 4.16 and 2.41, respectively. These superior thermal properties of the fluid (water, ethanol, etc.) enable the liquid metal droplets to be quickly cooled in the liquid environment. The metal objects with 3D structures can thus be rapidly prototyped. Besides, this new method prevents the liquid metal droplets from air oxidation which would guarantee the quality of the printed metal objects.

Table 2 Properties of water, ethanol and dry air at 100 kPa, 20 ℃

|  | Liquid phase cooling fluid | | Gas phase cooling fluid |
|---|---|---|---|
|  | Water | Ethanol | Dry air |
| Thermal conductivity ($\lambda$, W/(m·K)) | 0.597 | 0.24 | $2.59 \times 10^{-2}$ |
| $\lambda/\lambda_{air}$ | 23.05 | 9.27 | 1.00 |
| Density ($\rho$, kg/m$^3$) | 0.998 | 0.7893 | 1.205 |
| $\rho/\rho_{air}$ | 828.22 | 655.02 | 1.00 |
| Viscosity ($\eta$, Pa·s) | 0.001 | 0.0012 | $17.9 \times 10^{-6}$ |
| $\eta/\eta_{air}$ | 55.87 | 67.04 | 1.00 |
| Heat capacity ($c$, kJ/(kg·K)) | 4.1818 | 2.42 | 1.005 |
| $c/c_{air}$ | 4.16 | 2.41 | 1.00 |

When the liquid phase 3D printing method is administrated, several factors should be taken into account as they would significantly affect the final printing quality.

The first factor comes to the properties of the cooling fluid. The temperature of the cooling fluid will directly affect the printing effect. If this magnitude is set too high, the new droplet falling will melt together with the previous ones. The result would be that the formed structure is difficult to "grow". But if the temperature of the fluid is set too low, the droplet falling will be rapidly cooled and solidified since the heat transfer process is completed instantly. Some unique structures of 3D metal parts can be manufactured by controlling the velocity and direction of the cooling fluid. Besides, the viscosity of the cooling fluid will affect the falling time while the droplet and the cooling fluid density will affect the buoyancy of the droplet which has already been explained above.

Second, the injection speed and the needle diameter play important roles in the present fabrication process. It was found that both factors will affect the size of the droplet and the distance between two adjacent droplets. Such issue has been discussed before regarding to the deposition case in air.[23] With the increase of the injection speed, the diameter of the droplet will become smaller and the different droplets will be more closely-spaced. When the injection speed is increased to a certain value, the dripping will become jetting. Controlling of the dripping-to-jetting transition has important impact on the prototyping speed.

The droplet diameter is also affected by the size of the injection needle. Fig. 7 shows the statistical results of the size distribution of the droplets produced by using needles of varying sizes. The experiment was performed at room temperature with the ethanol cooling under the same injection speed. It can be seen that the injection droplet diameter becomes larger with the increase of the needle size. For those needles whose diameters are 0.16 mm, 0.34 mm, 0.51 mm and 0.84 mm, the major diameter range of the produced droplets are 30-40 mm, 60-80 mm, 80-100 mm and 100-180 mm, respectively.

There has been an approximate linear relationship between the droplet diameter and the interval time between adjacent droplets which is indicated in Fig. 8 (A). And the whole process when the interval time changes from 2.1 seconds to zero is shown in Fig. 8 (B). As the interval time becomes shorter, the droplets become smaller and smaller. The droplet shape changes from sphere to fusiformis, and gradually adjacent droplets join together. When the interval time finally becomes zero, the dripping-to-jetting transition is completed.

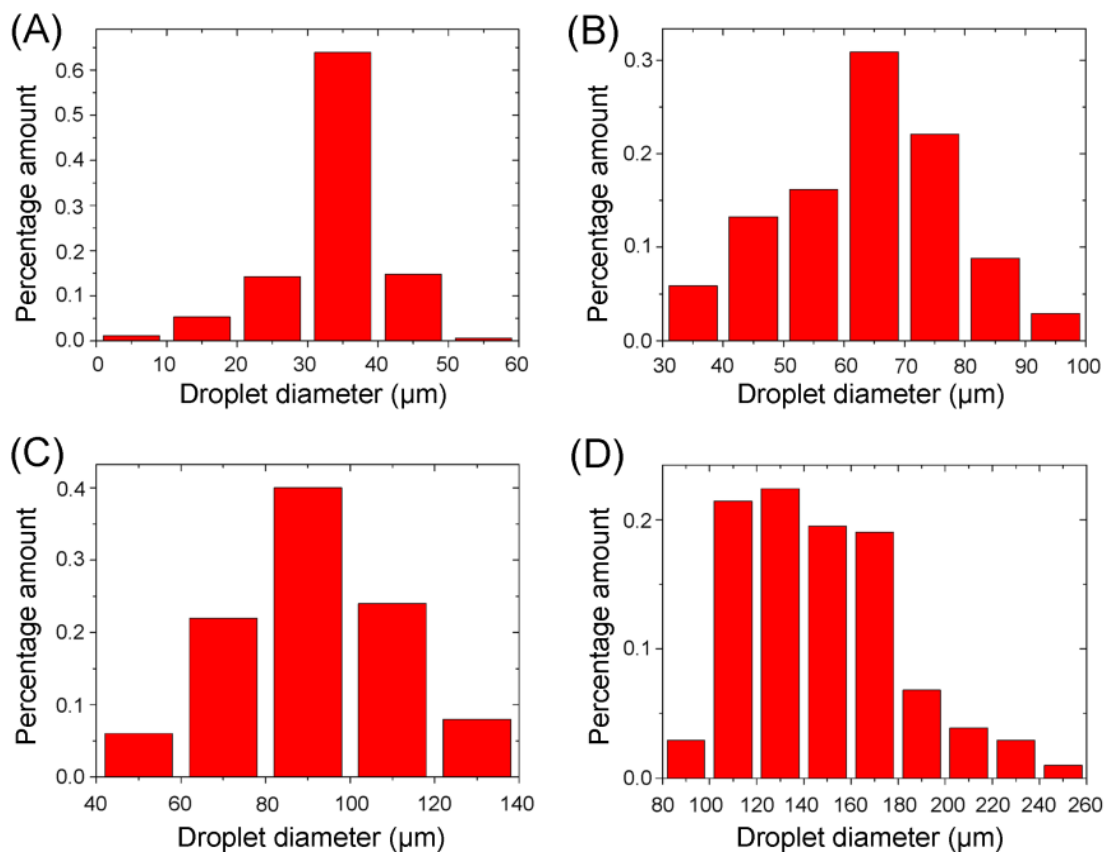

Figure 7 The size distribution of the droplets produced with syringe needles of 0.16 mm (A), 0.34 mm (B), 0.51 mm (C) and 0.84 mm (D) inner diameter, respectively.

Finally, the types and properties of the printing ink also dominate the fabrication process. In principle, all the low melting point (e.g. less than 300 ℃) metals can be selected as printing ink on condition that appropriate cooling liquid is available. The ink material can be gallium-, bismuth-, indium-based alloys, or even the mixture of these alloys and nanoparticles. Some properties of printing ink such as density, viscosity, surface tension, melting point, heat capacity, thermal conductivity can affect the shape of the 3D metal structure and the printing speed. The selection of the ink should be taken into consideration together with the cooling fluid type in the future.

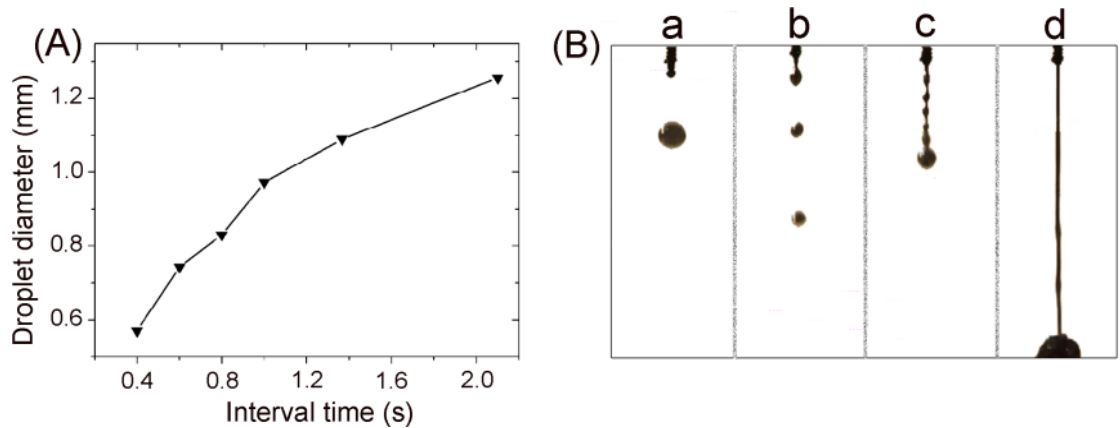

Figure 8 The relationship between the droplet diameter and the interval time between adjacent droplets. (A) Statistical results by using 0.26 mm inner diameter needle with ethanol cooling method; (B) The process (from a to d) when the interval time changes from 2.1 seconds (dripping) to zero (jetting).

Compared to the conventional metal prototyping method, liquid phase 3D printing method as proposed in this paper offers several evident advantages:

(1) Fast manufacturing speed. In the method of printing 3D metal objects in liquid phase, a fluid control mechanism is adopted and diversified 3D metal structures can be formed. In addition, the temperature field and flow field of the cooling fluid can be flexibly controlled. Through regulating the flow velocity and direction of the cooling fluid, some unique 3D metal structures can be realized, e.g. 3D rotating body.

(2) 3D electromechanical systems can be printed. Conductive liquid metal can be used in conjunction with nonmetal materials (e.g. plastic) to form 3D functional devices which include supporting structures and conductive devices. The combination of liquid phase 3D printing and conventional printing method can better meet all kinds of printing needs.

(3) The manufacturing energy consumption of metal components will be reduced. Due to introduction of low melting point liquid metal ink, the energy required when the solid ink melts is small. Therefore the difficulty of making metal components with this method is smaller than that using high melting point liquid metal via conventional method.

Then what will the liquid-phase 3D printer look like in the future? As a supplementary to the existing metal printing method, the liquid phase 3D printing method and system is expected to be further enhanced. The liquid metal printing ink can be gallium-based, bismuth-based alloys or other low melting point alloys. The printing process is carried out in a temperature balance space where the temperature should be higher than that of the liquid metal ink. To improve the accuracy and speed of the 3D printing, we would suggest adopt the combination between syringe pump array and syringe needle array. In such system, the syringe pump array is used to extract the liquid metal solution, while the syringe needle array is to inject the liquid metal ink into the cooling fluid. The injection needles can be replaced conveniently with other ones of different sizes to meet various printing demands. Designation and discretization of 3D models and controlling of the injection speed of each needle are completed through a computer-implemented process. The schematic diagram of injection needle array part is shown in Fig. 9. In this way, 3D metal objects are printed on the bottom of the constant temperature trough, in which the cooling fluid can be water, ethanol or others. Undoubtedly, the temperature of the cooling fluid should be set to a reasonable scale which is lower than that of the liquid metal ink. Besides, the flow velocity distribution of the cooling fluid should be controllable and some metal objects with unique structures can thus be directly printed out.

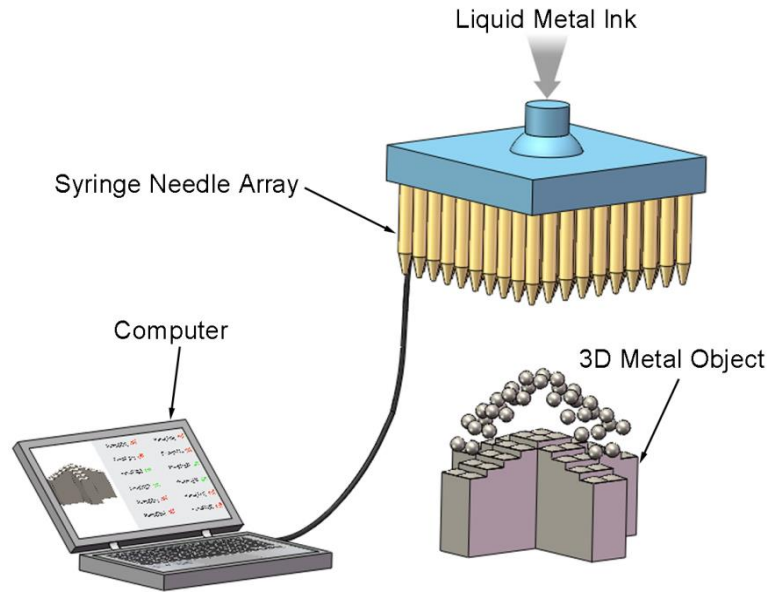

Figure 9 Injection needle array of future liquid phase 3D printer

**4. Conclusion**

In summary, we have established for the first time the liquid phase 3D printing method for directly making metal structures. Alloy inks with melting point above room temperature are identified as the printing inks. In a few cases, the liquid phase cooling method has shown merits over conventional air cooling method. With this strategy, 3D metal structures can be rapidly formed because of the high thermal conductivity and heat capacity of the liquid phase cooling fluid. Some important physical factors affecting the printing quality when using liquid phase method are also clarified. The size and falling velocity of the droplets can be flexibly controlled through modifying the syringe needle diameter, injecting speed, and the temperature, viscosity and surface tension of the printing ink and the cooling fluid, etc. The structures of the metal objects can be altered by changing the velocity and direction of the cooling fluid. Finally, a future liquid phase 3D printer was prospected. As a supplement to the existing 3D metal printing method, the liquid cooling printing also raised important fundamental as well as practical issues for solving in the coming time.